\documentstyle[preprint,aps,epsfig]{revtex}
\tightenlines
\begin{document}
\draft
\title{Spectral Statistics and Dynamical Localization: sharp transition 
       in a generalized Sinai billiard}
\author{Ulrich Gerland\cite{adr}}
\address{Max-Planck-Institut f\"ur Kernphysik,  
         69029 Heidelberg, Germany}
\date{\today}
\maketitle
\begin{abstract}
We consider a Sinai billiard where the usual hard disk scatterer 
is replaced by a repulsive potential with $V(r)\sim\lambda r^{-\alpha}$ 
close to the origin.
Using periodic orbit theory and numerical evidence we show that its 
spectral statistics tends to Poisson statistics for large energies 
when $\alpha<2$ and to Wigner-Dyson statistics when $\alpha>2$, while  
for $\alpha=2$ it is independent of energy, but 
depends on $\lambda$. We apply the approach of
Altshuler and Levitov [Phys. Rep. {\bf 288}, 487 (1997)] to show that 
the transition in the spectral statistics is accompanied by a 
dynamical localization-delocalization transition.
This behaviour is reminiscent of a metal-insulator transition in 
disordered electronic systems. 
\end{abstract}
%
%
The statistical distribution of quantum mechanical energy eigenvalues 
is of fundamental interest in diverse areas of physics such as 
condensed matter, atomic, and nuclear physics \cite{guhr98a}.
In the strict quasiclassical limit, where the de~Broglie wavelength is 
much smaller than all other lengthscales, the theory of quantum chaos 
relates the spectral statistics of a quantum system to the dynamics of 
its classical counterpart. While chaotic classical dynamics leads to 
Wigner-Dyson random matrix statistics \cite{bohigas}, 
integrable dynamics generically gives rise to Poisson statistics 
(uncorrelated eigenvalues) \cite{berry77a}. There is overwhelming 
evidence, both experimental and numerical, for these results 
\cite{houches89}.  
The spectral statistics of systems with mixed classical dynamics is 
expected to be described by superposing level sequences with Poisson 
and random matrix statistics, where the respective mean level spacing 
is determined from the size of the corresponding phase space volume 
with regular or chaotic dynamics \cite{berry84a}. 

However, when the de~Broglie wavelength is not the smallest lengthscale 
in the system, the spectral statistics is not solely determined by the 
classical dynamics. For example, the classical motion of an electron in 
a (three-dimensional) disordered system, such as a metal with substantial 
impurity scattering, can be regarded as completely chaotic. 
Nevertheless, upon variation of the disorder strength at a fixed energy 
(Fermi energy), the spectral statistics undergoes a sharp transition from 
Wigner-Dyson to Poisson statistics \cite{shklovskii}. 
Note that at the critical point the de~Broglie wavelength is of the same 
order of magnitude as the elastic mean free path, a classical 
lengthscale. The transition in the spectral statistics is 
accompanied by a transition from extended to localized eigenstates 
(Anderson metal-insulator transition) \cite{altshuler95a}. 
In the case of disordered systems one can therefore attribute the 
deviation of the spectral statistics from what one would expect on 
the basis of the classical dynamics to the quantum phenomenon of 
localization. 

A similar effect on the spectral statistics may be caused by 
{\em dynamical} localization \cite{casati79a,fishman82a}. 
Dynamical localization occurs for example in a circular billiard 
with a rough boundary, where in a certain range of parameters the quantum 
eigenstates are localized in angular momentum space despite that 
the classically chaotic dynamics leads to diffusive spreading of an 
initial angular momentum distribution \cite{frahm97}.
As a function of boundary roughness there is a crossover between localized 
and extended eigenstates, which takes place when the de~Broglie wavelength 
$\lambda_{\rm dB}$ is of the same order of magnitude as a classical 
lengthscale set by the roughness, $\langle(dR/d\theta)^2\rangle/R_0$, where 
$R(\theta)$ defines the rough circle in polar coordinates, 
$R_0=\langle R\rangle$, and the average is over the angle \cite{frahm97}. 
At the same time the level statistics changes smoothly from  
Poisson to Wigner-Dyson statistics \cite{frahm97,bogomolny99a}.

It is the purpose of the present article to show that a dynamical 
(non-random) system may also display a {\em sharp transition} in the 
spectral statistics. This is accompanied by a {\em dynamical} 
localization-delocalization transition of 
the eigenstates and is reminiscent of a metal-insulator transition in 
disordered systems. We thereby extend the list of analogies 
between dynamical and disordered systems \cite{altshuler95a}.

\begin{figure}[t] 
  \begin{center}
    \leavevmode
    \epsfig{figure=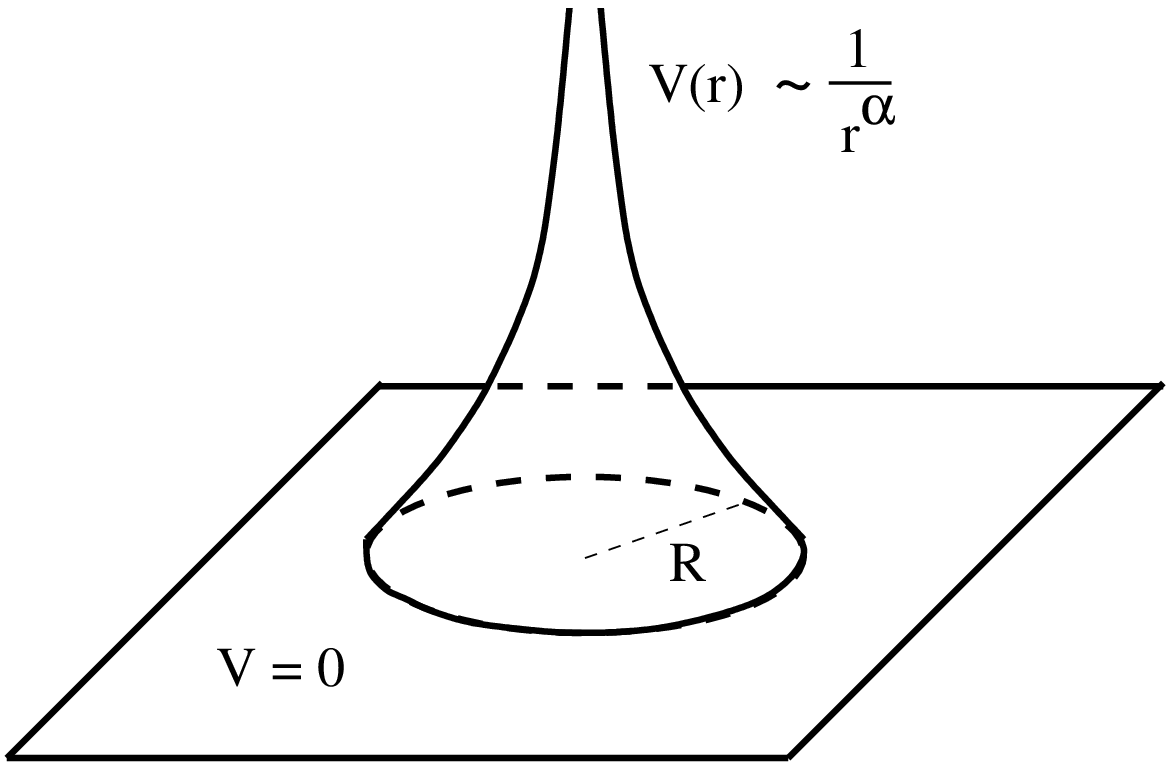,width=6.5cm}
  \end{center}
  \caption{Potential of the generalized Sinai billiard.}
  \label{fig1}
\end{figure}

We consider a generalization of the well-known Sinai billiard (SB). 
Sinai proved that the free motion of a classical particle
being specularly reflected from a disk of radius $R$ inside a 
square with periodic boundary 
conditions is completely chaotic 
\cite{sinai70a}. We modify the SB by replacing the disk with 
the scattering potential
\begin{equation}
  \label{potential}
  V(r)=\left\{\begin{array}{ll}
       \lambda\left(\left(\frac{R}{r}\right)^{\alpha}-1\right)\;,  
       & r<R \\ 0 \;, & r>R \end{array}\right. \;,
\end{equation}
where $\alpha$ and $\lambda$ are positive parameters (see also 
Fig.~\ref{fig1}).
Note that in the limit $\alpha\to\infty$ the SB is recovered. 
In a chaotic billiard, the replacement of a hard wall by a soft potential 
barrier generically leads to the formation of stable islands 
\cite{turaev98a}, however in the present case these are hardly visible in 
a Poincar\'e surface of section and are not relevant for the effect 
studied here. 

A similar system was considered by Altshuler and Levitov 
\cite{altshuler97a}. They focused on the properties of the eigenstates 
and proved the occurence of a dynamical localization-delocaliztion 
transition. In contrast, the emphasis in the present work is on a 
transition in the level statistics and on its semiclassical origin. 
The transition occurs as a function of $\alpha$ at $\alpha=2$,
irrespective of the value of $\lambda$. It is a sharp transition in the 
limit of large energy. Using periodic orbit theory we find that it is 
caused by a competition between regular 
(`bouncing ball'-type \cite{sieber93a}) and chaotic orbits. 
In the second part of this article we show that the approach of Altshuler 
and Levitov may be applied also to the present case. We thereby 
demonstrate that the transition in the spectral statistics is 
accompanied by a dynamical delocalization transition of the 
eigenfunctions.

For large energies, the qualitative quantum dynamics of the generalized 
Sinai billiard (GSB) is determined by the interplay of three lengthscales, 
(i) the de~Broglie wavelength $\lambda_{\rm dB}$, 
(ii) the radius $r_c=(E/\lambda+1)^{-1/\alpha}R$ of the 
classically forbidden area, and 
(iii) a typical length $\tilde{l}$ of an orbit before its direction is 
randomized by scattering from the potential.

When $\alpha>2$ the potential (\ref{potential}) effectively acts as a 
hard wall, since the radial wave function then vanishes like 
$\exp(-cr^{1-\alpha/2})$ with $c=2\sqrt{\lambda R^\alpha}/(\alpha-2)$ 
near the origin, i.e. faster than any power.
To estimate the behaviour of the spectral statistics, we approximate 
the potential (\ref{potential}) by a hard disk with the energy dependent 
radius $r_c$. According to the semiclassical theory \cite{berry85a}, the 
structure in the spectrum on the scale $\delta E$ is determined by periodic  
orbits of length $l=hv/\delta E$, where $h$ denotes Planck's constant and 
$v$ the velocity of the particle. For $l>\tilde{l}$ the chaotic orbits 
dominate and the spectral statistics will be random-matrix-like, while 
for $l<\tilde{l}$ the regular orbits dominate, leading to Poisson-like 
statistics. With $\varepsilon=\delta E/\Delta$ measuring the energy on 
the scale of the mean level spacing $\Delta$, we expect random-matrix-like
behaviour on energy scales $\varepsilon<\tilde{\varepsilon}$ and 
Poisson-like behaviour for $\varepsilon>\tilde{\varepsilon}$, where
$\tilde{\varepsilon}\sim A/\lambda_{\rm dB}\,\tilde{l}$ and $A$ 
denotes the area of the billiard. Now $\tilde{l}$ can be estimated by 
$\tilde{l}\sim A/r_c$, so that we find 
$\tilde{\varepsilon}\sim r_c/\lambda_{\rm dB}$. This ratio scales with 
energy as $r_c/\lambda_{\rm dB}\propto E^{(\alpha-2)/2\alpha}$.
Consequently, for $\alpha<2$ and increasing energy, $\tilde{\varepsilon}$ 
tends to zero, while $\tilde{\varepsilon}\to\infty$ for $\alpha>2$.
On this basis we expect a sharp transition between Poisson and random 
matrix statistics at $\alpha=2$ in the limit of large energy $E$.
Intuitively, the classically forbidden area becomes invisible to 
quantum mechanics for $\alpha<2$, while for $\alpha>2$ it becomes more 
and more sizeable.  

To see this explicitly, we apply Berry's periodic orbit theory of 
bilinear spectral statistics \cite{berry85a} to the SB with 
energy dependent disk radius $r_c$. The SB has two types of 
periodic orbits, those that never strike the disk and those that do. 
The former are marginally stable, occur in one-parameter families, and 
will be referred to as regular orbits, while the latter are unstable, 
isolated, and will be referred to as chaotic orbits.
In the following we determine the contribution of the regular orbits 
to the spectral form factor $K(\tau)$. 
Denoting the unfolded density of states by $d(\varepsilon)$, the form 
factor is related to the two-point correlation function 
\begin{equation}
  \label{R2}
  R(s)=\left\langle d\big(\varepsilon+\frac{s}{2}\big)\, 
  d\big(\varepsilon-\frac{s}{2}\big)\right\rangle_\varepsilon-1
\end{equation}
by Fourier transformation, 
$K(\tau)=\int_{-\infty}^{\infty}{\rm d}s\,e^{2\pi is\tau}R(s)$. 
One way to determine the contribution of the regular orbits to the density 
of states is to suitably modify the trace formula for the empty billiard.
To avoid degeneracies in the spectrum, we use a rectangular instead of
a quadratic billiard (sidelengths $a$ and $b$) and quasi-periodic
boundary conditions for the wavefunction, i.e. 
$\psi(x+a,y)=e^{i\phi_x}\psi(x,y)$ and analogously for the
$y$-direction \cite{symmetry}. 
The eigenvalues of the rectangle without the disk are 
$E_{jk}=(2\pi j+\phi_x)^2/a^2+(2\pi k+\phi_y)^2/b^2$ 
(here and below $\hbar^2/2m=1$). Applying 
the Poisson summation formula to the density of states of the empty 
rectangle and replacing the resulting Bessel function by its asymptotic 
form leads to a representation as a sum over periodic orbit families. 
Each family is specified by two integers, $m$ and $n$ 
(positive or negative), denoting the number of traversals across the 
billiard in $x$- and $y$-direction, respectively. Now including the 
disk obstructs the path of the orbits that violate the condition 
$2\,l_{mn}r_c(E)<A$, where $A=ab$ and $l_{mn}=[(ma)^2+(nb)^2]^{1/2}$ 
denotes the length of the orbits. The remaining families have to be 
weighted by the area they cover. The oscillatory part of the regular 
contribution to the unfolded density of states is then
\begin{equation}
  \label{regdens}
  \tilde{d}_{\rm reg}(\varepsilon)={\textstyle \sqrt{\frac{2}{\pi}}}
  \sum_{(m,n)'} f_{mn}(E)\,\sum_{j=1}^{\infty}
  \frac{\cos(jS_{mn}-\frac{\pi}{4})}{(j\,l_{mn})^{1/2}E^{1/4}}\;,
\end{equation}
where the first sum is over all primitive orbits and the second over 
repetitions. In eq.~(\ref{regdens}) we have introduced the actions 
$S_{mn}=l_{mn}\sqrt{E}+m\phi_x+n\phi_y$ and the orbit selection 
function $f_{mn}(E)=(1-2r_c(E)\,l_{mn}/A)\,\theta(A-2r_c(E)\,l_{mn})$, 
and set the mean level density to its large energy limit, so that  
$\varepsilon=AE/4\pi$.
This formula is a simple generalization of the expression for the 
usual SB given in Ref.~\cite{berry81a}. Substituting 
Eq.~(\ref{regdens}) into (\ref{R2}) and retaining only the 
diagonal terms in the double sum over periodic orbits (the diagonal 
approximation is justified by the energy average and known to be exact 
for regular systems \cite{berry85a}) yields the contribution of the 
regular orbits to the form factor,
\begin{equation}
  \label{KD}
  K_{\rm reg}(\tau)=
  \sum_{(m,n)'} f_{mn}^2(E)\,\sum_{j=1}^{\infty}
  \frac{\delta(\tau-j\,l_{mn}/A\sqrt{E})}{2\pi j\,l_{mn}\sqrt{E}}\;.
\end{equation}
This expression may be evaluated approximately by replacing the sums over 
primitive orbits and repetitions by a single sum over all orbits $(m,n)$.
In the large $E$ limit,
\begin{equation}
  \label{KDapprox}
  K_{\rm reg}(\tau)\sim \big(1-cE^{\frac{\alpha-2}{2\alpha}}\tau\big)^2\,
  \theta\big(1-cE^{\frac{\alpha-2}{2\alpha}}\tau\big)\;,
\end{equation}
with $c=2R\lambda^{1/\alpha}$. Eq.~(\ref{KDapprox}) shows that the 
contribution of the regular orbits to the form factor tends to zero 
for $\alpha>2$, while for $\alpha<2$ the Poisson limit $K(\tau)=1$ 
is attained. The contribution of the chaotic orbits assures that for 
$\alpha>2$ GOE statistics of random matrix theory is reached. To
see this explicitly would require to go beyond the diagonal
approximation. 

Next, we verify the above theoretical prediction of a transition in
the spectral statistics of the GSB numerically.
For $\alpha<2$, the eigenvalues of the GSB may be calculated by  
diagonalizing the Hamiltonian in the eigenbasis of the empty billiard.
However, for $\alpha\ge 2$ the matrix elements diverge and it is 
natural to work in the eigenbasis of the $1/r^\alpha$-potential 
\cite{divergence}.  
In this regime we chose to apply the Korringa-Kohn-Rostoker method, 
as described in Ref.~\cite{berry81a} for the case of the ordinary  
SB. The difference with respect to Ref.~\cite{berry81a} is 
that in the present case the scattering phase shifts of the potential 
have to be determined by solving the radial Schr\"odinger equation 
numerically, while they can be expressed in terms of Bessel functions 
for the usual SB. This makes the present case computationally 
more demanding. 

\begin{figure}[t] 
  \begin{center}
    \leavevmode
    \epsfig{figure=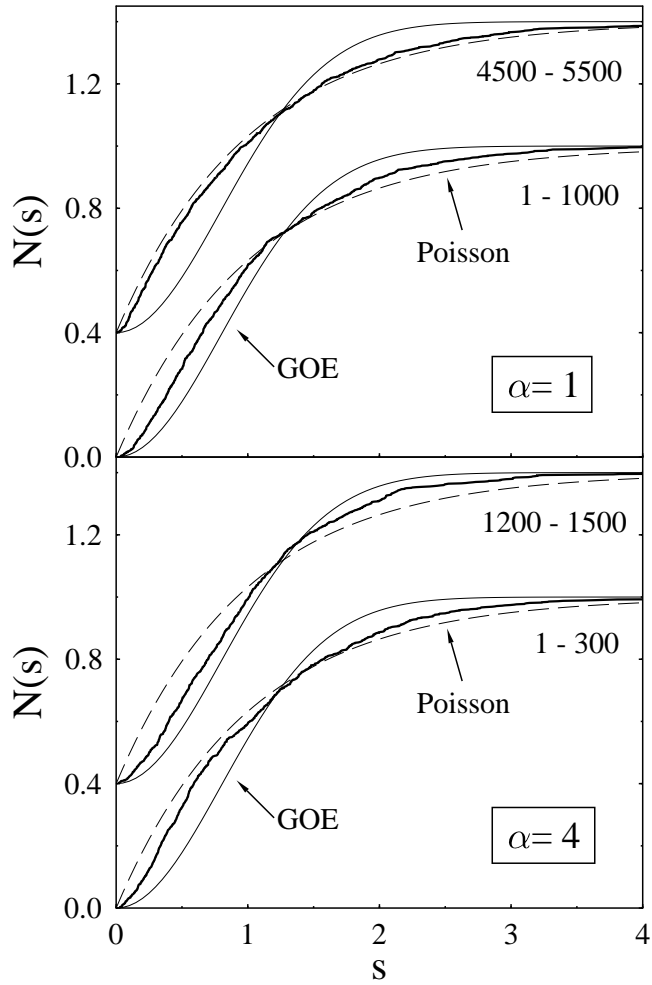}
  \end{center}
  \caption{Cumulative nearest-neighbour spacing distribution for the 
           generalized Sinai billiard. The energy levels that were used 
           are indicated below the curves (for $\alpha=4$ an average 
           over three different aspect ratios $a/b$ at constant area 
           $ab=4\pi$ was taken). With increasing energy a clear movement 
           towards Poisson statistics for $\alpha=1$ and towards GOE 
           statistics for $\alpha=4$ is visible.}
  \label{fig2}
\end{figure}

Fig.~\ref{fig2} shows the cumulative spacing distribution 
$N(s)=\int_0^s{\rm d}x\,P(x)$, with $P(s)$ denoting the 
nearest-neighbour spacing distribution, for $\alpha=1$ and 4 
(we used $\lambda=8$ for the former, $\lambda=0.01$ for the latter, and 
in both cases $R=1$ with a billiard of area $4\pi$). 
Each curve is displayed with two reference curves, $N(s)=1-e^{-s}$ 
for Poisson statistics and the Wigner surmise 
$N(s)=1-\exp(-\frac{\pi}{4}s^2)$ for the GOE. 
We observe a clear movement of $N(s)$ towards the Poisson curve with
increasing energy for $\alpha=1$ and towards the Wigner surmise 
for $\alpha=4$. Considering the relatively weak energy 
dependence of 
the spectral statistics, see Eq.~(\ref{KDapprox}), the limiting 
distributions are not expected to be reached within the available 
spectra. We verified that for $\alpha=2$ the 
spectral statistics is independent of energy in accordance with
Eq.~(\ref{KDapprox}). 


We turn to the properties of the eigenfunctions of the GSB. To keep the 
discussion general, we consider the GSB in $d\ge 2$ dimensions. 
Following Ref.~\cite{altshuler97a} we map the Schr\"odinger equation for 
the GSB on a localization problem. Applying Levitov's criterion of a 
diverging number of resonances \cite{levitov90a} we then show that 
$\alpha=2$ corresponds to a critical point associated with a 
delocalization transition.

Using periodic boundary conditions and a billiard 
of unit volume, the Schr\"odinger equation in momentum representation 
takes the form
\begin{equation}
  \label{schroeder}
  E_{\bf k}\,c_{\bf k} + \sum_{{\bf k'}\ne{\bf k}}
  V_{{\bf k}-{\bf k'}}c_{\bf k'} = E\,c_{\bf k}\;,
\end{equation}
where ${\bf k}$ denotes a site in the reciprocal lattice, 
$c_{\bf k}$ the Fourier coefficients of the wavefunction, $V_{\bf k}$ 
those of the potential, and $E_{\bf k}={\bf k}^2+V_{\bf 0}$.
Eq.~(\ref{schroeder}) may be interpreted as the Schr\"odinger equation 
of a particle on a lattice with on-site energies $E_{\bf k}$ and hopping 
amplitudes $V_{\bf k}$. Considering an eigenfunction with energy $E$, 
one finds \cite{altshuler97a} that its Fourier coefficients  
are non-vanishing only in an energy shell 
$E-\delta E<E_{\bf k}<E+\delta E$, where $\delta E$ is proportional to 
$\lambda$, which is assumed to be much smaller than $E$. Within this 
quasi-($d$$-$$1$)-dimensional shell, the on-site energies $E_{\bf k}$ are 
uniformly distributed quasi-random numbers. Since 
$V_{\bf k}\propto 1/k^{d-\alpha}$ for large $k$, one may interpret 
Eq.~(\ref{schroeder}) as an Anderson model with long-range power-law 
hopping within the energy shell \cite{altshuler97a}. 
A delocalization transition occurs when
the mean number of {\em resonances} per site diverges \cite{levitov90a}. 
A resonance is defined as a pair of sites $\bf k$,$\bf k'$ that fulfil 
the condition $|V_{{\bf k}-{\bf k'}}|>|E_{\bf k}-E_{\bf k'}|$.
If two sites are in resonance, the eigenstates of the corresponding 
two by two eigenvalue problem have amplitudes $c_{\bf k}$,$c_{\bf k'}$ 
that are comparable in magnitude. Informally, they can then be
considered as `linked' and an eigenstate can spread along this link.
If the mean number of resonances per site is infinite, 
there can be no localization, irrespective of the hopping strength. 

Noting that for $V_{{\bf k}-{\bf k'}}\ll\delta E$ \cite{resonances} 
the probability that the sites ${\bf k,k'}$ are in
resonance is simply $|V_{{\bf k}-{\bf k'}}|/\delta E$, the 
mean number of resonances ${\cal N}$ with a fixed site ${\bf k}$ 
can be estimated by summing the probability over the energy shell, 
\begin{equation}
  \label{Nres}
  {\cal N}=\frac{1}{\delta E}\sum_{{\bf k'}\,\in\,{\delta E}}
  |V_{{\bf k}-{\bf k'}}|\sim 
  \frac{1}{\sqrt{E}}\int\limits_0^{\sqrt{E}}{\rm d}k\,k^{d-2}\,V_k\;. 
\end{equation}
The sum is estimated by an integration over a flat region of radius 
$\sqrt{E}$ (we used $\Delta k=\delta E/2\sqrt{E}$ for the width of 
the energy shell and dropped an energy independent factor). 
Substituting the asymptotic form of $V_k$ for large $k$ 
into Eq.~(\ref{Nres}) we find that the mean number of resonances per 
site diverges in the limit of large energy when $\alpha>2$, independent 
of the dimension. In the case that was considered by Altshuler and 
Levitov \cite{altshuler97a} the same consideration leads to a 
critical $\alpha$ of one. For $\alpha$ below the critical value, the 
eigenstates are localized with a power-law tail due to direct hopping, 
independent of $\lambda$ when $d\le 3$ (because the effective lattice, 
i.e. the energy shell, is two-dimensional for $d=3$), and also for small 
$\lambda$ when $d>3$.

In summary, we have shown that the GSB displays a sharp transition in 
the spectral statistics, which is caused by a competition between 
chaotic and regular orbits. It is accompanied by a dynamical delocalization 
transition of the eigenstates. The investigation helps to clarify the 
relation between disordered systems and quantum chaos.

It is a pleasure to acknowledge useful discussions with E.~B.~Bogomolny, 
O.~Bohigas, A.~Mirlin, F.~v.~Oppen, C.~Schmit, U.~Smilansky and 
H.~A.~Weidenm\"uller.
%

%

\begin{thebibliography}{10}
\bibitem{divergence}
The divergence of the matrix elements is not related to the transition, 
since the former depends on the dimension $d$, while the transition 
occurs at $\alpha=2$ independent of $d\ge 2$.
\bibitem{resonances}
The finite number of sites that do not fulfil this condition may be
omitted without changing the resulting critical $\alpha$.
\end{thebibliography}
\end{document}